\documentclass{article}
\usepackage{graphicx} % Required for inserting images
\usepackage{geometry}[margins = 2cm]
\usepackage{natbib}
\usepackage{amsmath, amssymb}

\usepackage{booktabs}
\usepackage{longtable}
\newcommand*{\thead}[1]{%
\multicolumn{1}{c}{\bfseries\begin{tabular}{@{}c@{}}#1\end{tabular}}}
\usepackage{multirow}

\begin{document}

\title{Statistical Crime Linkage: Evaluating approaches within the Covenant for Using AI in Policing}

\author{
Nathan A. Judd\thanks{School of Mathematics, University of Birmingham, Edgbaston, Birmingham, B15 2TT, UK}
\and
Amy V. Tansell\footnotemark[1]
\and
Benjamin Costello\thanks{School of Psychology, University of Birmingham, Edgbaston, Birmingham, B15 2TT, UK}
\and
Liam Leonard\thanks{National Crime Agency, PO Box 8000, London, SE11 5EN, UK}
\and
Jessica Woodhams\footnotemark[2]
\and
Rowland G. Seymour\footnotemark[1]
}

\maketitle
\abstract{Linking crimes by modus operandi has long been employed as an effective tool for crime investigation. The standard statistical method that underpins statistical crime linkage has been logistic regression. The simplicity and interpretability of this approach has been seen as an advantage for law enforcement agencies using statistical crime linkage. In 2023, the National Police Chiefs' Council published the Covenant for Using Artificial Intelligence in Policing designed to guide the development of novel methods for use within policing. In this article, we investigate more statistical and machine learning methods that could underpin crime linkage models. We investigate a range of methods including regression-, sampling-, and machine learning-based techniques and evaluate them against the principles of Explainability and Transparency from the Covenant.  We investigate our methods on a new data set on romance fraud in the UK, where 361 victims of fraud reported the behaviours and characteristics of the suspects involved in their case. We propose a sensitive, Explainable, and Transparent machine learning model for crime linkage and demonstrate how this method could support crime linkage efforts by law enforcement agencies using a dataset of romance fraud with unknown linkage status. }

\section{Introduction}

Research indicates that the majority of crimes are committed by a minority of offenders \citep{Paulsen}. For example, a cohort study conducted in the United States, and consisting of $10,000$ people enrolled from birth, found that $6\%$ of offenders accounted for $52\%$ of delinquency, $71\%$ of murders, $73\%$ of rapes, $82\%$ of robberies, and $69\%$ percent of aggravated assaults \citep{wolfgang72}. Similarly, in the United Kingdom, \citet{Dodd04} found that $6-10\%$ of offenders accounted for more than half of the crime. Therefore, linking crimes by suspects can allow for the pooling of evidence from multiple crimes and support police officers in apprehending and convicting suspects. Linking crimes can also aid a more efficient approach to investigations whereby a series of crimes is investigated collectively rather than each crime individually.

%Furthermore, there is a considerable literature on decision support tools for crime linkage, e.g., \citep{TONKIN2017}, which can be based upon the aforementioned model choices.

The standard statistical crime linkage method is underpinned by logistic regression. This method has performed well in many applications, e.g., \citet{Tonkin12}. Two advantages of logistic regression are its simplicity and interpretability. More recently, other underpinning methods have been considered, such as Naive Bayes \citep{Crime_Linkage_Bayes_Factor}, Bayesian models, and Iterative Classification Trees \citep{TONKIN2017}. A natural next step is to consider whether machine learning (ML) could offer improved predictive performance in crime linkage.

Machine learning methods have been deployed to support professionals in everyday tasks by improving performance, optimising workflow, and speeding up procedures. For example, machine learning-based decision support tools have been developed to support: dentists in deciding what treatment patients with periodontal disease require based on patient records \citep{SVM_health}; teachers in deciding how to deliver online learning sessions that respond to students' attention levels \citep{ML_education}; and governments in assisting job-seekers to gain employment based on employment data \citep{Berman2024}. In policing, many artificial intelligence (AI) technologies have already been adopted, for example, drones, Automatic Number Plate Recognition (ANPR), and facial recognition \citep{facial_recognition_policing}. Furthermore, there are increasing numbers of AI and ML decision support tools being developed to support decision-makers in law enforcement, e.g., \citep{Shenvi_Smith_23}, as well as the justice system more widely, e.g., \citep{Amy_CEG}.

It is important that law enforcement are open about their use of machine learning and AI technologies and that their tools are developed responsibly and robustly within the public interest. To support the responsible development and use of AI and machine learning within policing, the UK National Police Chiefs' Council released guidelines in the form of the 2023 `Covenant for Using AI in Policing' (henceforth referred to as the \textit{Covenant}) \cite{AIinPolicing}. The Covenant outlines six principles to guide the use of AI within UK policing, namely, \textit{Lawful}, \textit{Transparent}, \textit{Explainable}, \textit{Responsible}, \textit{Accountable}, and \textit{Robust}. These straightforward principles have both statistical and organisational implications. The principles most pertinent to statistical methodological development are Explainable and Transparent, where Explainable refers to the ability of an algorithm to explain its output, both in terms of how it was reached and what it can say, and Transparency refers to the ability of third party experts to examine, understand, and test these methods.

In this article, we conduct a methodological comparison of a suite of regression-, sampling-, and machine learning-based techniques with respect to both predictive performance metrics and the principles of Explainable and Transparent in the Covenant. We demonstrate these methods on a novel dataset comprising cases of romance fraud.

This paper is structured as follows: Section \ref{sec data} describes a novel dataset comprising cases of romance fraud. Section \ref{sec methods} contains subsections on related work, model evaluation criteria, and the methods considered. In particular, Section \ref{sec lit rev} reviews pertinent literature on similar types of data either qualitatively (i.e., crime linkage) or statistically (i.e., binary classification problems). Sections \ref{sec reg methods}-\ref{sec ml methods} describes an array of methods, both standard and machine learning-based, that can be used to link crimes and discusses how they fit into the Covenant. Section \ref{sec results} presents the results of each method having been deployed on the romance fraud dataset with known linkage status. Appropriate statistical metrics and graphical representations are used to support the comparison and inferences. Section \ref{sec best} presents the most suitable method, and discusses its composition and predictive performance. In Section \ref{sec unsolved}, this chosen method is then deployed on another dataset comprising cases of romance fraud where linkage status is unknown (rather than known). To conclude, Section \ref{sec discussion} summarises the research undertaken in this article and its implications for law enforcement agencies and governmental institutions.

\subsection{An motivating dataset}\label{sec data}

The aim of this article is to develop a sensitive machine learning-based crime linkage model that fits into the Covenant. To facilitate this, we use a new dataset comprising cases of romance fraud.

%This is the first time that machine learning methods have been developed for statistical crime linkage pertaining to an online crime. Furthermore, the model developed is a bespoke blend of modern machine learning methods and statistical crime linkage.

Romance fraud occurs when an individual is deceived for financial gain by someone with whom the victim believes themselves to be in a romantic relationship \citep{Romance_Fraud_Definition}. Our dataset includes extreme cases whereby deception continues over many years, the victim is defrauded of hundreds of thousands of pounds, and/or the victim supports the trafficking of an offender to a new country. Romance fraud is predominantly an online crime, and in our dataset the fraud was found to be initiated exclusively online. The data was provided by Action Fraud, the UK's national reporting centre for fraud and cybercrime. Reports came in two forms: self-reports submitted to Action Fraud via an online reporting tool; and case notes from calls to Action Fraud where the call handler summarised and then submitted the victim’s report on their behalf.

These text-based reports were cleaned and anonymised before being imported into the NVivo qualitative data analysis software (Lumivero, Denver, CO, U.S.) for analysis. NVivo was used to create a binary dataset from the reports by manually coding the absence ($0$) or presence ($1$) of features within each report. We constructed a codebook comprising $400$ features of romance fraud based on information obtained from a review of the literature, psychological expertise within the team, and discussions with law enforcement. The codebook was iteratively and inductively developed based on the content of the data. Two researchers independently coded the data, and an inter-rater reliability test was carried out to ensure both researchers agreed on the coding. Once coding had been completed, researchers met to discuss the coding, any discrepancies across the data, and to agree on the final coding. Ethical approval was given by the University of Birmingham Science, Technology, and Mathematics Ethics Committee (ERN\_1075-Apr2023).

One report was found to be a duplicate and another to contain two separate cases of fraud implicating two different suspects. After adjusting for these, $361$ unique cases of romance fraud were identified in the data, of which $61$ were solved and $300$ unsolved. Each case comprised $400$ features, excluding linkage status, before data pre-processing and cleaning. The \textit{response variable} (or dependent variable) was linkage status, i.e., whether two cases of fraud against two different victims were committed by a common suspect (\textit{linked}) or by two different suspects (\textit{unlinked}). The data on linkage status were provided by Action Fraud in the form of a $61\times 2$ matrix with columns indicating the victim case number and the suspects' randomised ID. The \textit{features} (or independent variables) range from demographic information (e.g., gender) to information pertaining to the crime itself (e.g., whether the `foot-in-the-door' method of extracting money from the victim was used \citep{Woodhams06}). The latter type of entries were interpreted by the researchers using the aforementioned procedure. In general, features were grouped into one of the following eight categories of characteristics: \textit{Victim}, \textit{Suspect}, \textit{Temporal}, \textit{Initial Approach}, \textit{Follow-On Contact}, \textit{Maintenance}, \textit{Extortion}, and \textit{Closure}.

%For the purposes of this article, we split the $361$ cases into a solved dataset comprising $61$ cases and an unsolved dataset comprising $300$ cases. We trained a model on the solved cases and used a testing subset to ev

To adopt the standard approach to statistical crime linkage, we transformed the raw data into a pairwise dataset so that each row represented a pair of cases. The columns consisted of binary vectors and for any given row (pair of cases) and column (feature), an entry was either a $1$ if the features of the cases matched or a $0$ if they did not match. For example, letting $P$ be the number of features in the dataset, the $n$th row would be a vector of length $P$, taking values in $\{0,1\}^P$, representing the similarity between cases with randomised IDs $1$ and $n$. We also created a vector encoding the linkage status of the pairs. A pair of cases are linked (encoded as $1$) if they share the same offender and are unlinked if they do not (encoded as $0$). As such, an ideal model of linkage status (i.e., with linkage status as the dependent variable) and rows of similarity in features (as the independent variables) should map similarity in modus operandi to linkage status.

After the pairwise dataset was created, we had to address a high level of homogeneity and missingness present in some features. In particular, the level was high for demographic features such as ethnicity, religion, and sexuality (for both the victim and suspect), and low for the psychological descriptors of the relationship. All homogeneous columns, i.e., all columns consisting entirely of $0$s or $1$s (excluding missing entries), were removed. As such, $150$ features, excluding linkage status, remained. 

Regarding missingness, no pair of victim cases had a complete set of features, and so, the issue had to be addressed, given that listwise deletion would remove all the data. One approach was to merge some sets of feature. For example, type of follow-on-communication was reduced from seven categories (email, text message, instant messenger, mobile phone, call, video call, and face-to-face) to three (text-based, call-based, and face-to-face). Features with greater than $10$\% missingness were centred around demographic-, geographic-, or platform-based information. Features describing psychological characteristics of communication in the initial, maintenance, extortion, and closure phases of the relationships were less than $10$\% missingness. Some variables believed to be important and with a moderate amount of missingness, whilst also believed to be missing completely at random (MCAR) or missing at random (MAR), were retained. For example, the gender of the suspect was missing in $37.5$\% of cases. However, it is reasonable to believe this was either MCAR or MAR, and if it was MAR, it was likely related to the other observed variables. Hence, multiple imputation was a favourable option as compared to listwise deletion, which would result in a considerable loss of cases and information. Similarly, \lq\lq Reasons For Money\rq\rq was missing in $27.5$\% of cases, but due to the richness of categories and their importance in describing the extortion method of the crime, it is undesirable to use listwise deletion. This variable could be assumed MCAR if the reason was impersonal and forgotten or omitted due to perceived unimportance. In the cases where a reason was missing, and this was a deliberate omission, the observed psychological descriptors variables would likely predict its category and hence MAR would be assumed. Outside of these, features with greater than $25$\% missingness were removed from the dataset. Tests for MCAR and MAR could not be implemented due to co-linearity in the feature set, and so the remaining features were assumed to be either MCAR or MAR. Missing values were imputed using multiple imputation with chained random forest models available through the \texttt{missRanger} R package \citep{Stekhoven11}. After conducting this missingness procedure, $51$ features, excluding linkage status, remained, with some categories reduced in size and the Temporal category removed entirely. Most notably, the reduced Initial Approach category had a lower level of co-linearity which is required for most modelling approaches.% However, the multiple imputation procedure itself may have introduced some co-linearity in the imputed variables, but due to the reduced prevalence of missingness, the co-linearity remained low.

\section{Methodology}\label{sec methods}

In this article, we are interested in linking crimes. There is a long history of crime linkage in the criminal psychology literature, see e.g., \citep{Woodhams2019}. Novel statistical approaches to crime linkage have also been undertaken, e.g., \citep{Crime_Linkage_Bayes_Factor,TONKIN2017}.

From a statistical perspective, crime linkage is a clustering problem and, using pairwise data, it is adequately formulated as a binary classification problem. For such problems, a confusion matrix is a good visual aid for model evaluation. Alongside a confusion matrix, and its associated metrics such as Sensitivity, Specificity, and Precision, the Area Under the Receiver Operating Characteristic Curve (AUROCC) and the Area Under the Precision-Recall curve (AUPRC) can also be useful model evaluation metrics.

For the purpose of this article, \textit{sensitivity} and \textit{specificity} will be primary metrics for model evaluation. Law enforcement often strive to minimise false negatives (FNs), that is, to not miss any linked pairs of victims. As such, sensitivity is valued more than specificity. In spite of this, a model that is $100\%$ sensitive would be useless unless it had a positive (i.e., non-zero) specificity. As such, all methods were optimised via grid search with respect to the sensitivity metric. Given ties in parametrisations of a method, grid search was implemented once again to optimise for specificity.

In addition, a method- and parameter-specific threshold to divide predicted linkage status can be implemented to accommodate for the internal logic of non-probabilistic methods, account for the depression of predicted probabilities caused by the class imbalance, and encode the defined costs of FN and false positive (FP) classifications. The most-sensitive reparametrisation for each method were used to draw a range of model summary metrics to facilitate a wider evaluation of predictive performance. A final method was then chosen based on its sensitivity and specificity, as well as its Explainability and Transparency.

%Any machine learning methods in the crime linkage context should be evaluated not only by their predictive performance but by their Explainability and Transparency, as suggested in the Covenant.

After selecting the most suitable method and parametrisation, we predicted linkages in a dataset where the linkage status was unknown. These predicted linkages encode a network representing clusters of victims sharing the same suspects. This output is valued in policing and government policy as it estimates how many suspects there are in a dataset. As such, the methodologies considered in this article are of interest to law enforcement agencies and governments.

% HERE

\begin{figure}[!h]
    \centering
    \includegraphics[width=1\linewidth]{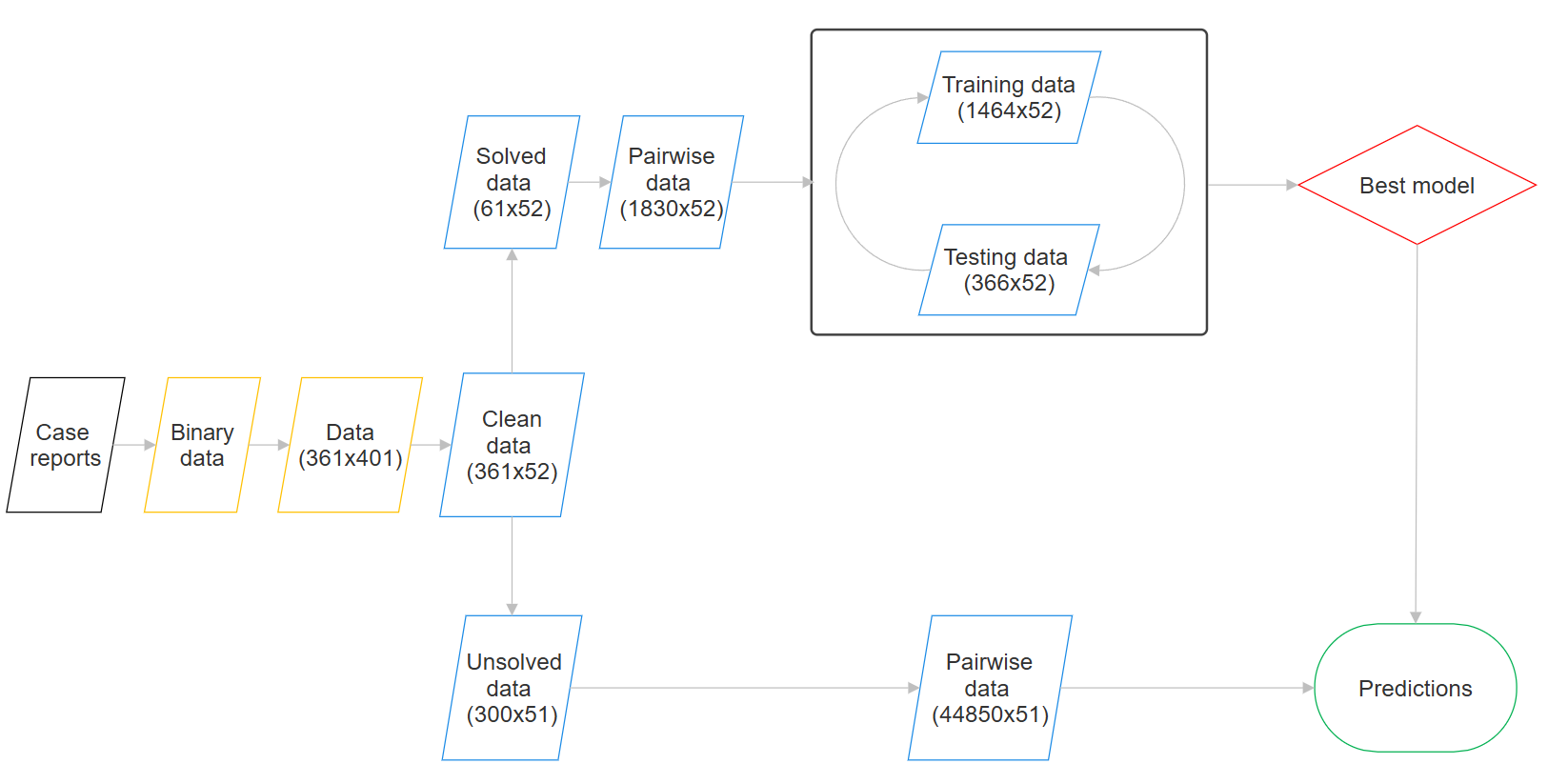}
    \caption{A workflow diagram of our methodology.}
    \label{fig:workflow}
\end{figure}

Figure \ref{fig:workflow} depicts the data workflow and model-selection strategy undertaken in this article. In particular, we note that the top flow represents the model building procedure and the bottom flow represents the predictions on a dataset with unknown linkage.

\subsection{Related work}\label{sec lit rev}

We now proceed to review some pertinent literature relating to the statistical problem and the data application. Linking crimes by modus operandi is a growing approach to criminal investigations. The standard method is based on logistic regression and it has been applied successfully. For example, \citet{MO_using_LR_ROC} achieved an accuracy of $67.6$\% and an AUROCC of $0.81$ when linking burglaries. Logistic regression is often favoured due to its simplicity and interpretability. However, other statistical approaches have also been successful. For example, in \citet{Crime_Linkage_Bayes_Factor}, a naive Bayes classifier for pairwise case-linkage achieved an accuracy of 82\% with a FP rate of 5\% on a dataset of $10,670$ breaking and entering crimes. A recent comparison of regression-, tree-, and Bayesian-based methods was conducted on a large dataset of cases of sexual assault spanning five countries \citep{TONKIN2017}. The authors discovered that one Bayesian method achieved an AUROCC of $91\%$ whilst logistic regression only achieved $87\%$; although we emphasise that there are differences between our motivating data and theirs. The findings in \citet{TONKIN2017} suggest that predictive performance in crime linkage could be improved by considering non-standard methods. However, these methods may have costs and limitations. This is why we use the Covenant to guide methodological development.

%More recently, machine learning has been applied to spatio-temporal modelling of crime frequencies/levels \citep{ZHANG2022}.

The previous literature on crime linkage is arguably qualitatively different to the dataset of romance fraud cases considered in this article. For example, $19\%$ of the $4,681$ solved cases in \citep{Crime_Linkage_Bayes_Factor} were linked. As such, the class imbalance is around $5$ times smaller and the dataset $15$ times bigger than our dataset on romance fraud (with $\sim 4.86\%$ of the pairs linked). Furthermore, the current dataset pertains almost-exclusively to online crime, where geographic distance, an important feature in crime linkage, has a different structure and quality to, e.g., burglaries. In addition, the location of the suspect was often missing from the dataset. Similarly, temporal information was generally inaccurate or unreported. The combination of these three differences (data size, minority class size, and absence of some key features) presents a difficult predictive task. The pairwise approach, with a binary feature space, may also result in existing models being underpowered. The use of rebalancing methods and machine learning may help to amplify and extract the signal from these data.

Machine learning methods have recently been applied to criminal data, e.g., spatio-temporal modelling of crime-counts in \citep{ZHANG2022}. Furthermore, machine learning methods have been successful in binary classification problems. It is therefore natural to consider machine learning models for crime linkage. However, for these models to be used in the public domain, predictive performance should not come at the expense of Explainability and Transparency.

There is also a significant literature on classification methods for unbalanced datasets. We refer the reader to \citep{review_of_classification_unbalanced} for a review. The most common methods are under- and over-sampling (of the majority and minority classes, respectively) before fitting any model \citep{review_of_classification_unbalanced}. Two popular developments include synthetic over-sampling technique (SMOTE) \citep{Chawla_2002} and Random Over-Sampling Examples (ROSE) \citep{ROSE, Menardi_Torelli_2014}, the latter of which will be detailed in Section \ref{sec methods}.

In this article, we explore three groups of methods: regression-, sampling-, and machine learning-based methods. Sampling-based methods include under- and over-sampling (of the majority and minority classes, respectively) and combinations thereof \citep{review_of_classification_unbalanced}. However, these methods are not adapted to, nor have been tested against, a dataset with predominantly binary features, as is the case in this work. Regression-based methods such as logistic, Lasso, Ridge, and Elastic net regression are weak classifiers when deployed on imbalanced datasets. Owing to the objective functions associated with their fitting procedures, these methods have the tendency to overfit to the minority class. In spite of this, regression-based methods fit in well with the Covenant due to their Explainability (e.g., they have interpretable parameters). That said, machine learning methods have the potential to capture more complex relationships in data, but their compliance with the Covenant may prevent them from being used. As such, we predict that a combination of methods may be desirable.

%The work in this paper differs from the existing literature in that (1) the crimes considered were exclusively initiated online, which represents a challenging but important domain for statisticians to develop methods for [citation needed], (2) the models considered are \lq\lq black-box\rq\rq, which have been sparsely used in crime linkage [citation needed], and, (3) the AI/machine learning methods are developed explicitly in conjunction with the 2023 AI in Policing Covenant. We also discuss evaluation metrics and provide a graphical representation of predictions to support decision-makers.

\subsection{Model evaluation}

In this paper, we compare a suite of methods against the principles in the Covenant as well as a broad range of common model summary metrics. To evaluate the predictive performance of each method, we use the $61$ solved cases of romance fraud, i.e., the cases of romance fraud for which there has been a conviction or law enforcement have a high degree of certainty on the offender. We then fit the models to a training set ($80\%$) and evaluate their performance against a testing set ($20\%$). This approach prevented methods from overfitting to the training data.

We evaluate the Explainability of a method by the reach, identifiability, and interpretation of the output produced. For example, we are interested in the following properties: the availability probabilities or pseudo-probabilities; small changes in inputs have a one-to-one correspondence with small changes in outputs; and the mapping from input to output of the method is understandable and interpretable.

We evaluate the Transparency of a method by its ease of reproduction and its openness to unit testing. The Covenant says that law enforcement should document the algorithms employed and that methods should be open to adversarial third-party testing. These principles enable an understanding of how the model works in the absence of a clear mechanism from input to output. Therefore, the Transparency of a method addresses a key criticism of machine learning methods. Ideally, for public openness, a third party should be able to: reproduce the method and verify that they can produce the same outputs when using the same inputs; and check that there are no unforeseen or undesirable biases. A method could fail to be reproducible for reasons such as the use of private or outdated software packages, poor description or documentation, or the absence of seed control. If a method were not reproducible then law enforcement would have failed in their duty to inform the public of their use of advanced algorithms. Furthermore, the public would not be reassured that there were no undesirable behaviours or biases in the machines outputs.

With regards to evaluating model performance, we will provide a wide range of summary metrics evaluated on an independent testing dataset. In particular, we present the sensitivity (SE), specificity (SP), precision (P), harmonic mean of the sensitivity and specificity (H), and accuracy (ACC). These metrics are central to binary classification problems.

%****** davis and goadrich, 2006 / drymmond and holte, 2006 / he and garcia 2009 / Perkins \& Schisterman *******

For the purpose of our application, the sensitivity metric is of greater importance than either accuracy or the area under the Receiver Operating Characteristic (ROC) curve. This is because the cost of missing a link is much higher than the cost of misclassifying an unlinked pair as linked. This motivation is common in other domains such as fraud detection and disease screening.

It has been suggested that the precision-recall curve is more appropriate than the ROC curve for both classification problems on imbalanced datasets \citep{Saito15} and crime linkage \citep{AUC_Crime_Linkage}. However, the precision-recall curve can be problematic and unrepresentative for certain non-probabilistic machine learning models, such as those considered in this article. Therefore, the AUROCC will be used to present predictive performance as a metric of models for varying preferences between sensitivity and specificity.

\subsection{Regression-based methods}\label{sec reg methods}

The standard method in the statistical crime linkage literature is to model linkage using a logistic regression model with either summaries of groups of features \citep{BURRELL24} or, more recently, a single similarity summary. We will refer to these models as logistic regression $6$ (LR6) and logistic regression $1$ (LR1), where the integer refers to the number of non-intercept parameters. In addition, temporal and geographic distance or classifications can be added as independent variables to the predictor, separate from the group summary components. This helps avoid an excessive number of features and reduces collinearity within the feature set. In particular, the Jaccard coefficient is chosen to represent the similarity between two vectors of features \citep{Woodhams2019}, where the Jaccard coefficient $J: \mathbb{R}^P \times \mathbb{R}^P \longrightarrow [0,1]$, is defined by
\begin{equation} \label{eqn jaccard coef}
J(A,B) := \frac{|A \cap B|}{|A \cup B|} \qquad \forall\, A, B \in \{0,1\}^P
\end{equation}
where $P$ is the number of independent variables (features).

In the context of crime linkage, the standard model evaluation method for logistic regression has been leave-one-out cross-validation (LOOCV). In our dataset, the LR1 and LR6 models achieve indistinguishable predictive performance metrics for the different evaluation methods such as LOOCV, grid-search and stepwise selection. As such, grid-search is implemented for the suite of methods considered in this work to enable a smoother comparison with other classes of methods.

%$J$ represents the similarity between the features of a pair and has long been used in the crime linkage literature \citep{Woodhams2019}.

We now consider Elastic Net (EN) regression model, which is a form of penalised regression. EN models are setup exactly the same as generalised linear models (GLMs), but the objective function in the fitting procedure is modified to penalize model complexity by reducing both the number of parameters and the magnitude of the fitted coefficients. In particular, this is achieved by adding $|\boldsymbol{\beta}|$ and $|\boldsymbol{\beta}|^2$ to the objective function in the fitting procedure. Two extra parameters, given by $\alpha$ and $\lambda$, control the magnitude and distribution of the penalty to both objectives. When $\alpha=0$, EN models reduce to Ridge regression models and, when $\alpha=1$, they reduce to least absolute shrinkage and selection operator (LASSO) regression models. In Ridge regression, regression coefficients are encouraged to approach zero, but they cannot actually reach it. As such, Ridge regression is a variance-reducing method. LASSO regression is a standard method in applied statistics for data where the number of parameters is much larger than the number of observations; for example, see \citep{RICH2020}. As such, LASSO is a covariate-selection method.

GLMs are more Explainable than penalised regression methods because uncertainty quantification is available in the form of $p$-values. Penalised regression methods fit biased coefficient estimates and, thus, a distribution of the estimate around the true value cannot be formed to quantify uncertainty. Nonetheless, coefficients returned by penalised regression are interpretable and map small changes in features to small changes in the predicted probability of being linked. Thus, non-penalised and penalised regression methods are adequately and moderately Explainable, respectively.

Regression-based methods are adequately Transparent. Both methods have the same minimisation-based fitting procedure, but with different objective functions. Regression-based methods are well studied and have proliferated into all statistical applications. Such models are easy to reproduce and standard libraries for these models also exist on a variety of open source platforms.

\subsection{Sampling-based methods}

Random over-sampling (ROS) amplifies the minority class within the dataset as a whole by re-sampling with replacement and returning a superset of the original data. In particular, Random Over-Sampling Examples (ROSE) creates an interpolated space from the feature-space of the minority class, which avoids merely increasing the multiplicity of the observed minority class (a cause of overfitting) \citep{Menardi_Torelli_2014}. It includes a parameter $p$ which determines the proportion of the minority class within the ROSE re-sampled data. ROSE is available in the \texttt{ROSE} R package \citep{ROSE}.

Synthetic minority oversampling technology (SMOTE) is similar to ROSE, except that the additional minority class samples are uniformly generated from the line connecting a minority class example to one of its nearest neighbours (in the feature space of the minority class). ROSE generally outperforms SMOTE due to the enlarged feature space used to generate the additional samples \citep{Menardi_Torelli_2014}. Note that there is not much attention given to binary (or categorical) state-spaces within the literature on oversampling methods and, as such, there is opportunity for future methodological development.

Sampling-based methods are adequately Explainable. The re-sampling parameter is identifiable (given a known seed) and leads to stable and predictable changes from input to output. Sampling-based methods are adequately Transparent. The methodologies are well-documented and implementable by third-party testers. Existing libraries are also available from trusted, high-quality platforms.

\subsection{K-nearest neighbours filter}

The $k$-nearest neighbours filter (KF) method was introduced in \citep{kf_kang}. In applications with two classes, the algorithm swaps class labels if the $k$-nearest neighbours of a data point are all members of the other class. In the context of imbalanced class problems, we implement this method to re-label minority class data points to majority class labels and not the other way round. This is because we are concerned with the signal-to-noise ratio within the minority class and outliers in the minority class are more disruptive compared to in the majority class (where the sample size is far greater). The definition of a neighbour is canonically that of the $k$-nearest neighbours algorithm, i.e., the Euclidean distance. However, owing to the high-dimensional, predominantly-binary feature space presented by the romance fraud dataset, we chose to adapt this method by replacing use of the Euclidean distance with the Jaccard coefficient. This summary statistic represents similarity distances between cases in our dataset. The KF algorithm was principally designed to reduce noise and amplify signals in the minority class, although it has also been seen to protect against misclassification. In spite of this, the method has the potential to bias the resultant subset of the original data. The KF method is parametrised by a natural number $k$ which governs the number of majority-class neighbours required to swap a minority label with a majority label.

In this article, we calculate the $k$-nearest neighbours of a pair by determining the Jaccard coefficient displayed in \eqref{eqn jaccard coef}. This avoids computing distances in a predominantly-binary feature space.

The parameter of the KF algorithm is identifiable and leads to predictable changes in outputs, although small increases in the parameter can lead to significant changes in the output (minority and majority class labels in the outputted dataset). The mechanism of the algorithm is simple to understand when compared to other methods and the usefulness of the output is proportional to the algorithm's purpose in our methodology. (The internal role of the Jaccard coefficient might also add interpretability \cite{Woodhams2019}.) As such, the KF method is adequately Explainable. The algorithm's simplicity and one-to-one mapping between inputs and outputs makes it easily tested by third-parties and so, the KF method is adequately Transparent.

\subsection{Machine learning-based methods}\label{sec ml methods}

We now consider the Explainability and Transparency criteria for a selection of machine learning methods. In particular, we consider, Support Vector Machines (SVMs), Random Forests (RFs), and Extreme Gradient Boosting (XGBoost).

Shapley Additive Explanations (SHAP) \citep{Lundberg2017AUA} has been shown to offer post-hoc values for the importance of features in any predictive model. It is often useful for machine learning methods where there are no internal parameters, such as regression coefficients, offering a similar explanation. The method computes a weighted average contribution for each feature in the model by comparing model predictions with the feature and without the feature. A kernel SHAP method was developed (and deployed in R via the \texttt{kernelshap} package) which does not assume independence between features \citep{AAS2021}. This can be used in lieu of in-built Explainability, equivalent to regression-based methods.

\subsubsection{Support vector machines}

An SVM is a \emph{supervised learning} algorithm. Its original and simplest form projects the $P$-dimensional feature space \emph{linearly} onto a $Q>P$ dimensional space before an optimisation procedure identifies the optimal hyperplane able to linearly separate the (labelled) training data. Minority classes can be assigned weights in unbalanced datasets and some non-linear projections are supported (e.g., polynomial, radial, and sigmoidial). To enable classification by the algorithm, each data point lying on the boundary of the classifier requires a support vector. Linear projection leads to a minimisation procedure similar to those in quadratic programming, and, remarkably, non-linear projections come at no additional computation cost \citep{vapnik92}. 

SVMs can be computationally expensive. In the case where the number of support vectors approaches the number of data points, $n$, a constrained optimisation problem is obtained. All SVMs have a cost parameter, $c$, with the radial (non-linear) projection introducing an additional parameter, $\gamma$. A larger cost encourages the SVM to overfit to the (training) data by fitting a decision boundary as tight as possible around the data, whilst a smaller cost prefers to maximise the decision boundary, and thereby encourage a more generalisable model.

By default, SVMs only output binary predictions. There exist methods to produce `probabilities', but these are unsupported by theory and unreliable in practice.. As such, the Explainability of an SVMs output is limited. In spite of this, in combination with the post-hoc Shapley values method for computing feature contributions in predictions, SVMs become moderately Explainable.

SVMs are intuitive, but in high-dimensional settings reproducing the code could be difficult (although not impossible) for third-party testers. Unit testing is also available, given numerous third-party packages available on quality platforms. Shapley values can be computationally expensive to compute. Overall SVMs are moderately Transparent with or without SHAP.

\subsubsection{Random forests}

RFs \citep{random_forests} are constructed by bagging (bootstrap aggregating) a collection of decision trees, with each decision tree trained on a bootstrapped sample (of the original training dataset) comprising a randomly-selected subset of features. This method is generalisable and avoids overfitting, but it is less interpretable compared to regression-based models. For instance, the user could investigate the weights for each decision tree and subsequently investigate a subset of important decision trees, but this can be time-consuming and inaccurate. The main parameters of RFs are the number of trees, $N_t$, and the number of features randomly subsampled to train each tree, $M$.

It is notable that a collection of decision trees is harder to interpret and comprehend when compared to a single decision tree. However, summaries of the `most-representative' tree can be made, where, for example, decision thresholds and edge weights can be derived from the RF model via averaging. A (single) decision tree is intuitive and is similar to regression-based methods in interpretability. As such, RFs are an adequately Explainable method. RFs use standard methodologies such as bootstrap aggregating and decision trees, enabling reproduction by third parties. Further, third-party packages also exist for unit testing. Overall, RFs are adequately Transparent.

\subsubsection{XGBoost}

XGBoost \citep{xgboost} is a C++ library supporting a machine learning algorithm based on gradient boosting of collections of regression trees, seminally introduced in \citep{Friedman2001}. In comparison to RFs, where trees are trained in parallel, XGBoost is a boosting method where trees are trained sequentially. The optimisation, speed, and stability of the XGBoost algorithm is what sets it apart from other gradient boosting methods. In particular, XGBoost uses advanced regularisation techniques to suppress weights and prevent overfitting, whilst its cache-coding and parallel-core abilities speed up the C++ library. The XGBoost library allows for many parameters, including, but not limited to, the scale of the learning rate, $\eta$, the number of rounds, $N_r$, and the depth of each tree, $D$. The Explainability and Transparency of this method is comparable to RFs.

XGBoost results in a single decision tree iteratively tuned to the data. A decision tree is intuitive and easy to read and interpret, making it easier to interpret than a collection of trees like a RF. However, how the method arrives at its final decision tree, i.e., how the inputs lead to outputs, is not as clear as for RFs. As such, XGBoost is adequately Explainable, comparable to RFs but with nuanced differences.

XGBoost uses a lot of advanced algorithms and computing resources, making reproduction difficult for third parties outside of testing existing packages. As a result, XGBoost is less Transparent than RFs and can be considered moderately Transparent.

A summary of each method and their compatibility with the statistically relevant principles in the Covenant is displayed in Table \ref{table_methods_principles}.

\begin{table}
\caption{The methods and their level of Explainability and Transparency. (Coding: L=Limited, M=Moderate, A=Adequate.)}

    \label{table_methods_principles}
    \centering
    \resizebox{\textwidth}{!}{ % Resize table to fit within text width
        \begin{tabular}{c|cccccccc}
            \hline
            Method group   &  Regression & Penalised Regression & Sampling & KF & SVM & SVM (w/ SHAP) & RF & XGBoost \\
            \hline
            Explainability   & A & M &  A & A & L & M  & A & A \\
            Transparency  &  A & A & A & A & M & M & A & M \\
        \end{tabular}
    }

\end{table}

\section{Results}\label{sec results}

In this section, we fit each method described in Section \ref{sec methods} to the subset of the romance fraud data with known linkages. Recall that there were $61$ victims, each with a known suspect who committed the crime. From these, a pairwise dataset of $1830$ pairs was constructed, with features taking the value $1$, if both cases in a pair had the same feature value, and $0$ otherwise. The training set consisted of $80\%$ of these $1830$ pairs, which were selected at random, with the remaining pairs making up the testing set. On the testing set, all methods were optimised for sensitivity using grid search over their respective parameter spaces. All computations used a seed of one. The grids used are presented in Table 2 in the Supplementary Material. All methods remained stable with respect to small perturbations in their parameters. In fact, the methods generally produced the best sensitivity for multiple parameterisations and subsequently underwent optimisation with respect to specificity.

We use a method-parameter-specific threshold for binary classification as opposed to the standard threshold of $50\%$. Other crime linkage work has used Youden's J statistic, which balances sensitivity and specificity. As law enforcement wish to prioritise sensitivity, we describe a method where sensitivity is preferred, and we can define the extent to which it is preferred. We introduce a cost parameter (as well as the ratio of the minority and majority class prevalences), which is interpreted as the relative ratio of the cost of FNs relative to FPs \citep{Perkins2006}, and can be readily implemented via the \texttt{pROC} package in R \citep{Robin2011}. See the supplementary material for more details on this method. Our approach allows law enforcement agencies to set the cost ratio parameter depending on their organisational approach to risk. To demonstrate in our data set, we estimate the cost of a false negative and false positive linked crimes. The cost of a false positive link is the cost incurred to the law enforcement agency for investigating a pair of cases that are not be committed by the same suspect. The UK Home Office estimate the total cost excluding overheads in 2007/08 for investigating fraud cases was $£1,040$ \citep{Alan}. For a false negative, we estimate the cost of a suspect not being identified. From our training data set, suspects commit a small number of offences  (median three offences), and as a conservative estimate we assume that suspects commit one more offence. We set the economic cost of a this to be the median amount of money stolen in our data set, which is £$1,690$. The cost parameter is the ratio of these values, i.e. $1.625$ ($\frac{1690}{1040}$), and the weight parameter in the minimisation procedure is $r=(1/1.625)\times (0.9514/0.0486)\footnote{This fraction represents the ratio of the prevalences of the majority and minority classes where we recall that $0.0486$ is the prevalence of linked crimes.}=12.05$. The class imbalance is so large that despite the cost ratio, the overall weight $r$ still favours specificity over sensitivity; nevertheless, exploratory analysis indicates that these thresholds achieve greater sensitivity than the standard $50\%$ threshold. We acknowledge this method is crude and does not take into account any of the wider economic or emotional impacts. However, this approach allows law enforcement agencies to direct their model search towards their own cost-benefit perception and tolerance for risk.

\begin{table}[h]
\caption{The methods, and their performance metrics (\%), for cost-model-specific thresholds. (Legend: SE = sensitivity, SP = specificity, P = precision, HM = harmonic mean, ACC = accuracy.)}
\label{tab: model comparison}
\centering
\resizebox{\textwidth}{!}{
\begin{tabular}{ c|cccccc } 
Method  & SE & SP & P & HM & ACC & AUROCC   \\
\hline
    \thead{Logistic regression 1 \\ $\boldsymbol{\beta}=( -2.96, -0.0833  )$} & 27.78 & 87.64 & 10.42 & 15.15 & 84.7 & 54.41
 \\
 \hline
 \thead{Logistic regression 6 \\ $\boldsymbol{\beta}=(-3.24, 0.599, -0.423, 0.12, 0.0656, -0.206, 0.297
)$} & 11.11 & 89.08 & 5 & 6.9 & 85.25 & 41.08 \\
 \hline
 \thead{EN \\ $\lambda=0.0115 ,\alpha=a$} & 5.56 & 99.71 & 50 & 10 & 95.08 & 37.31 \\
 \hline
 \thead{KF-EN  \\  $k=3,\lambda=0.0121,\alpha=0.1$} & 5.56 & 99.71 & 50 & 10 & 95.08 & 37.15\\
  \hline
 \thead{ ROSE-EN \\ $p=0.31,\lambda=0.00805 ,\alpha=0.1$} &72.22 & 42.24 & 6.07 & 11.21 & 43.72 & 42.43
  \\
 \hline
 \thead{KF-ROSE-EN \\ $k=2,p=0.43,\lambda=0.006,\alpha=0.5$} &  77.78 & 27.01 & 5.22 & 9.79 & 29.51 & 40.68 \\
 \hline
 \thead{SVM \\ $c=0.014$}  & 22.22 & 91.95 & 12.5 & 16 & 88.5 & 44.7 \\
 \hline
 \thead{ KF-SVM  \\ $k=1, c=0.003$} & 22.22 & 87.07 & 8.16 & 11.94 & 83.88 & 51.18 \\
 \hline
 \thead{ROSE-SVM \\ $p=0.34;c=0.01$} & 77.78 & 47.99 & 7.18 & 13.15 & 49.45 & 57.42
\\ 
 \hline
 \thead{KF-ROSE-SVM \\ $k=1
 ;p=0.29;c=0.014$} & \textbf{83.33} & 45.4 & 7.32 & 13.45 & 47.27 & 49.55 \\ 
\hline
 \thead{Random Forest Regression \\ $N_t=20,M=7$} &  66.67 & 44.54 & 5.85 & 10.76 & 45.63 &  54.1 \\
 \hline
 \thead{XGBoost \\ $\eta=0.8,N_r=300,D=6$} & 72.22 & 55.17 & 7.69 & 13.9 & 56.01 & 58.75 \\
 \hline
 
\end{tabular}
}
\end{table}

Table \ref{tab: model comparison} presents the predictive performance metrics associated with the methods from Section \ref{sec methods} that were fitted and tested on the solved cases of romance fraud.

Since law enforcement aim to minimise false negatives (i.e., not miss any linked cases), our principal interest lies with comparing the sensitivities of the methods. The standard methodology for crime linkage is represented by LR1 and LR6.   The unforeseen difference in sensitivity between these two methods (as seen in Table \ref{tab: model comparison}) is likely a result of the lower responsiveness to changes in the threshold of the LR6 model induced by the fitting of seven regression coefficients as opposed to two (as in LR1). As a result, LR6 overfits to the minority class and produces a better overall ROC curve and is a better model for balanced problems in terms of perceived costs and class sizes. However, for the specific class imbalance and FN cost perception considered in this article, LR1 outperforms LR6 due to the flexibility provided by having fewer parameters. In spite of this, $27.78\%$ sensitivity is still very low and our machine learning-based methods are seen to improve on this by around $240-300\%$ (as seen in Table \ref{tab: model comparison}). The most notable models are KF-ROSE-EN, KF-ROSE-SVM, and ROSE-SVM, with the performance of  ROSE-EN and XGBoost being notable. Table \ref{tab: model comparison} reveals KF-ROSE-SVM to be the most sensitive method, achieving a sensitivity of $83.33\%$ and retaining a competitive specificity of $45.4\%$. Aided by Shapley values, KF-ROSE-SVM is Explainable and Transparent and is recommended as the most suitable method.

\subsection{Most suitable method}\label{sec best}

The final model is a trained SVM with $c=0.014$ and $1313$ support vectors, of which $890$ and $423$ belong to the majority and minority classes, respectively. Given that the re-sampled class sizes were $29\%$ and $71\%$, respectively, the distribution of support vectors closely resemble that of the number of data points in each of the classes. This suggests that the model is allocating a roughly proportional amount of resources to minimising FPs and FNs.

\begin{table}[h]
    
    \caption{Confusion matrix for the optimal method when testing the dataset.}
    \label{tab: confusion matrix}
    \begin{tabular}{c  r | c  c | c}
        \multirow{4}{*}{\textbf{Actual outcome}} 
        & & \multicolumn{2}{c}{\textbf{Predicted outcome}} & \\
        & & \textbf{0} & \textbf{1} & \textbf{Total} \\
        & \textbf{0} & 158 & 190 & 348 \\
        & \textbf{1} & 3 & 15 & 18 \\
        & \textbf{Total} & 161 & 205 & 366
    \end{tabular}
\end{table}

\begin{figure}[h]
    \centering
    \caption{Network of pairs of victims in the testing dataset, with coloured edges representing whether the LR1 (left) and KF-ROSE-SVM (right) models identified links or not.}
    \label{fig: test_pred}
    \includegraphics[width=1\linewidth]{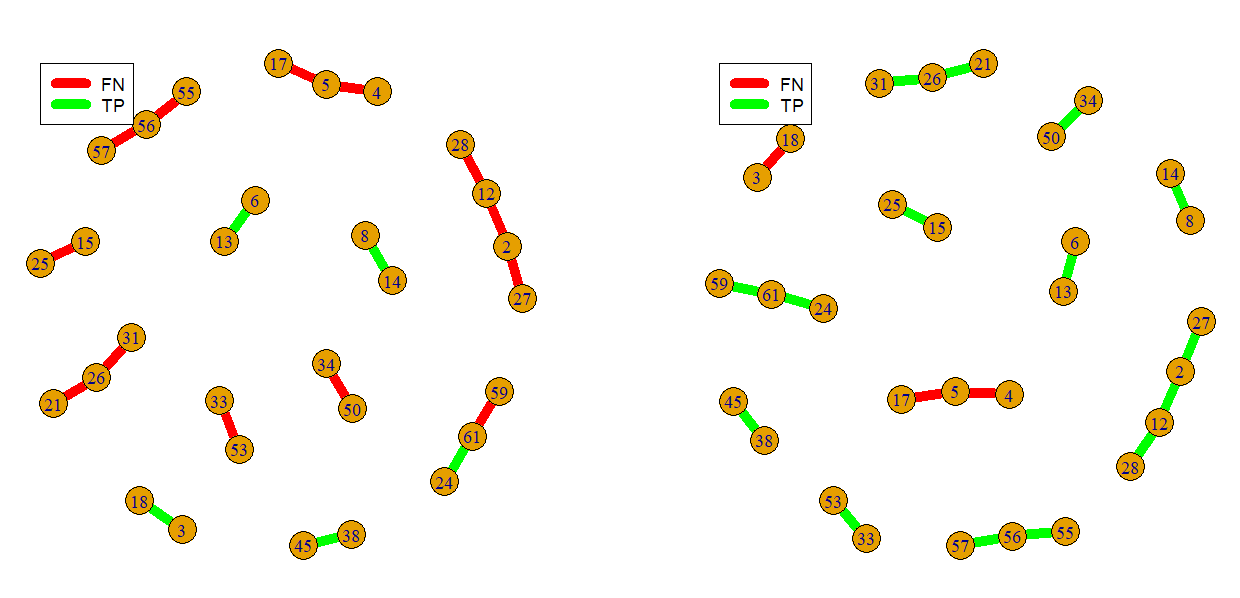}
    \vspace{-1cm}

\end{figure}

From Table \ref{tab: confusion matrix} we observe that the method predicted $158$ \lq phantom\rq linkages but missed three. Figure \ref{fig: test_pred} presents the linkage structure of the testing dataset and reveals the links identified and missed by the trained SVM. The model missed three links which form two clusters, and so, we can conclude that the model identified $10$ suspects but failed to capture two.
\\

Figure \ref{fig: test_pred} captures the victims and linkages in the testing dataset. The left and right plots have different colour codes representing the performance of the LR1 and KF-ROSE-SVM methods, respectively. Combining data from Figure \ref{fig: test_pred} and Table \ref{tab: confusion matrix}, we can estimate any savings in resources and labour inflicted on law enforcement by employing KF-ROSE-SVM on a dataset of similar size and character to our training data, when compared to LR1. KF-ROSE-SVM identified $10$ of the $12$ suspects when compared to LR1, which identified five. As a counterfactual argument, the KF-ROSE-SVM method identified seven more suspects than the LR1 method. As before, we assume that each suspect goes on to commit one more offence, so using the KF-ROSE-SVM method instead of LR1 would mean £11,830 not being stolen from victims in this data set.

\begin{table}[ht]
\centering
\caption{SHAP feature contributions in descending order of absolute magnitude}
\label{tab: shap results}
\begin{tabular}{|p{4.5cm}|r|p{5.5cm}|r|}
\hline
\textbf{Feature} & \textbf{Weight} $\times 10^3$ & \textbf{Feature} & \textbf{Weight} $\times 10^3$ \\
\hline
Total Amount & 0.42 & Male Suspect & 0.42 \\
Victim Special & -0.32 & The Perfect Partner & -0.29 \\
Technological Expertise & -0.27 & Crazy Making Abuse & 0.24 \\
Third Party Involvement & -0.23 & Foot-In-The-Door & 0.23 \\
Suspect Sends Photos/Videos & 0.23 & Threatens Victim & -0.18 \\
Female Suspect & -0.16 & Female Victim & -0.16 \\
Emotional Manipulation & 0.14 & Discouraging Disclosure & 0.12 \\
Threats/Ultimatums & 0.11 & Money Mule & -0.11 \\
Communication Type: Text & -0.10 & Victim Suggests Alt. Payment Help & 0.08 \\
Male Victim & -0.06 & Suspect Placed Victim in Power & 0.06 \\
Vulnerable Suspect & -0.06 & Travel to Suggested Destination & -0.06 \\
Escalating Crisis & 0.05 & Suspect Impersonates Someone Else & -0.04 \\
Sad Story & -0.04 & Frequent Small Payments & -0.03 \\
Repeat Victim & 0.03 & Role Escalation & -0.03 \\
Reasons For Money Category & -0.02 & Victim Degradation & 0.02 \\
Vagueness & -0.02 & Ceases All Contact & 0.02 \\
Working Towards Shared Goal & 0.02 & Grammatical Errors / Poor English & 0.01 \\
Deletes Profile & -0.01 & Suspect Opens Bank Account for Victim & -0.01 \\
Job Offer & 0.01 & Multiple Suspect & -0.01 \\
Communication Type: F2F & 0.01 & Sign for Goods on Behalf of Suspect & 0.01 \\
Victim Sends Personal Details & -0.00 & Promises of Sex & 0.00 \\
Love Bombing & -0.00 & Financial Gain & -0.00 \\
Repeated Requests for Funds & 0.00 & Communication Type: Call & 0.00 \\
Excuses & 0.00 & One Off Request for Funds & -0.00 \\
Victim is the Same & -0.00 & Victim Sends Intimate Images/Videos & 0.00 \\
Persistent Communication & -0.00 & & \\
\hline
\end{tabular}
\end{table}

Table \ref{tab: shap results} presents the Shapley values for each of the $51$ features in the dataset. From this, we can interpret the relative importance of each feature in the model's predictions of linkage status. For example, the magnitude of the weights corresponds to their importance in predictions, whereas the sign corresponds to whether the feature's presence is associated with a higher or lower probability of linkage. In particular, \lq\lq Persistent Communication\rq\rq ~ and \lq\lq Victim Sends Intimate Images/Videos\rq\rq ~had the smallest magnitudes, and, as such, represent the least-influential features in the model. As such, we can conclude that these features are equally common in linked and unlinked cases of romance fraud. Important features in the model included: 1) the total amount stolen, (2) Male Suspect, (3), Victim Special, (4), The Perfect Partner, (5) crazy-making abuse, (6) the foot-in-the-door method, and, (7), a suspect sending photographs and/or videos. Note that (3) and (4) were associated with a lower probability of linkage, whereas (1), (2),  (5), (6) and (7) were associated with a higher probability of linkage. We conclude that these seven features are unequally distributed between the linked and unlinked pairs of cases of romance fraud. Consult Table 3 in the Supplementary Material for definitions of the the features.

\subsection{Cases with unknown linkage status}\label{sec unsolved}

In this section, we use the trained model from Section \ref{sec best} to predict linkages between pairs of cases of romance fraud where linkage status is unknown.

The KF-ROSE-SVM model predicted $17,729$ links out of a possible $44,850$ when using the cost-specific cut-off from the KF-ROSE-SVM evaluation on the testing data. It is reassuring to find that the cut-off lies within the tight range of predicted probabilities generated by the SVM model, despite being on a far larger dataset.

We are also able to use this method to better understand the prevalence of offenders. Three, $84$, and $143$ suspects were predicted upon applying the Louvain cluster method with resolution one, two, and three, respectively. (Other community detection algorithms could also be used.) The resolution can be tuned based on the requirements of specific law enforcement agencies, and this choice should depend on the expected ratio of victims to suspects. Based on the solved data comprised of $61$ victims and $17$ suspects, we chose a resolution of two. This choice corresponded to $84$ suspects and $300$ victims and better suited the ratio of suspects to victims found in the solved data. Table \ref{tab: num of victims} presents the distribution of suspects and the number of victims. Of the $84$ suspects, $54$ had a single victim. The more prolific suspects had $23$, $40$, and $54$ victims each. 

\begin{table}
\centering
        \caption{The predicted suspect-victim distribution.}
    \begin{tabular}{|c|c|c|c|c|c|c|c|c|c|c|c|c|c|}
    \hline
        Number of suspects & 54 & 3 & 9 & 5 & 2 & 2 & 3 & 1 & 1 & 1 & 1 & 1 & 1 \\
        \hline
    Number of victims & 1 & 2 & 3 & 4 & 5 & 6 & 8 & 9 & 10 & 11 & 23 & 40 & 54 \\
    \hline
    \end{tabular}

    \label{tab: num of victims}
\end{table}

\begin{figure}[h]
    \centering
    \caption{Three clusters of victims (left, middle and right) corresponding to three suspects.}
    \label{fig: three clusters}
    \includegraphics[width=1\linewidth]{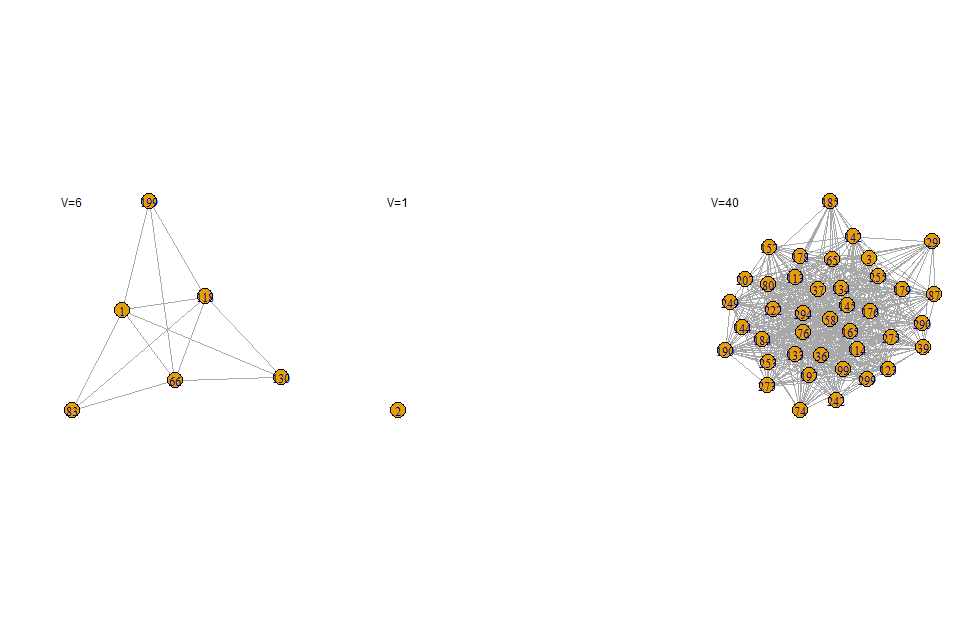}
    \vspace{-1cm}

\end{figure}

Figure \ref{fig: three clusters} presents the three clusters identified by the Louvain method with resolution $2$. Where Table \ref{tab: num of victims} is useful for both governmental and law enforcement agencies, as an estimate on prevalence, Figure \ref{fig: three clusters} could be useful for law enforcement agencies in their specific information-gathering and crime-modelling tasks. This is because law enforcement agencies can link new cases to solved cases or pool evidence from multiple victims to support investigations and prosecutions of suspects.

\section{Discussion}\label{sec discussion}
In this article, we discussed and evaluated a diverse suite of methods against predictive performance metrics as well as the principles of Explainability and Transparency in The Covenant. We determined that standard methods in the crime linkage literature had poor sensitivity (an important metric in policing) as a result of the extreme class imbalance in the dataset on romance fraud. We considered a range of methods that provided a significantly greater sensitivity, with the $k$-nearest neighbours filtered ROSE-rebalanced Support Vector Machine method yielding the largest sensitivity (as well as a strong specificity). KF-ROSE-SVM is Explainable and Transparent, especially when supporting the method post-hoc with Shapley values. As such, this model was chosen to predict the linkage network within the $300$ cases of romance fraud with unknown linkage status. In doing so, we found the method was able to estimate the number of suspects and the number of victims associated with each suspect. The proportion of suspects to victims was similar to that in the testing dataset, which is reassuring.

This method could be helpful to both law enforcement and governmental agencies. In particular, the method could be used to estimate the prevalence of suspects in emerging or difficult-to-gauge areas of crime. Moreover, this method may also be of interest to criminal psychologists who are interested in the relationship between features and linkage status. Future work could involve deploying these methods on larger datasets and on datasets containing different categories of crime.

In conclusion, machine learning offers significant improvements in the predictive performance of crime linkage models of cases of romance fraud. The methods considered allow law enforcement to prioritise Explainability and Transparency or predictive performance, such as sensitivity and accuracy. Given that a minority of offenders commit the majority of crime, the evidence provided in this article suggests that, by using a KF-ROSE-SVM crime linkage approach, law enforcement may enable an improved identification of suspects and a larger evidence pool from investigations. This could lead to better law and order outcomes as well as reducing the public expense of these crimes. The approach taken in this article could be used by the government to estimate the prevalence of criminals in different domains of crime, and help allocate resources for emerging crimes or crimes for which the prevalence is difficult to gauge. We have also discussed how improvements in predictive performance can be achieved without deviating from the Ethical use of AI, as outlined in the 2023 UK Covenant for Using AI in Policing.

\section{Acknowledgments}

This work was supported by a UKRI Future Leaders Fellowship [MR/X034992/1] and the QR Policy Support Fund. We thank the National Crime Agency for their support in accessing the romance fraud data set.

\bibliographystyle{abbrvnat}
\bibliography{main}
\end{document}